\documentclass{PoS}

\usepackage{epsf}
\usepackage{amssymb}
\usepackage{amsmath}
\usepackage{amsfonts}
\usepackage{cite}
\usepackage[small]{caption2}

\usepackage[T1]{fontenc}
\usepackage{psfrag,epsfig,graphicx,graphics}

\newcommand{\scalingm}{2.2cm}

\newcommand{\scaling}{2.98cm}

\newcommand{\scalingl}{4cm}

\newcommand{\scalingA}{2.3cm}
\newcommand{\scalingB}{2.8cm}
\newcommand{\scalingC}{3.2cm}
\newcommand{\scalingD}{2.9cm}

\newcommand{\be}{\begin{equation}}
\newcommand{\beq}{\begin{equation}}
\newcommand{\ee}{\end{equation}}

\newcommand{\bea}{\begin{eqnarray}}
\newcommand{\eea}{\end{eqnarray}}

\def\slashchar#1{\setbox0=\hbox{$#1$}
   \dimen0=\wd0
   \setbox1=\hbox{/} \dimen1=\wd1
   \ifdim\dimen0>\dimen1
      \rlap{\hbox to \dimen0{\hfil/\hfil}}
      #1
   \else
      \rlap{\hbox to \dimen1{\hfil$#1$\hfil}}
      /gdatdafinal2.tex
   \fi}

\title{Production of $ \rho^0_L-$meson pair in $\gamma^* \gamma^* $
collisions}

\ShortTitle{Production of $ \rho-$meson pair in $\gamma^* \gamma^* $
collisions}

\author{B.\ Pire$^a$, \speaker{M.\ Segond}\,$^b$,
L.\ Szymanowski$^{b,c,d}$ and
S.\ Wallon$^b$ \\
        ${}^a$\,CPHT,\,Ecole Polytechnique, CNRS, Palaiseau, France \\
        ${}^b$\,LPT, Universite Paris-Sud, Orsay, France \\
        ${}^c$\,Universit\'e  de Li\`ege, Belgium \\
	${}^d$\,So{\l}tan Institute for Nuclear Studies, Warsaw, Poland \\
        E-mail: \email{Mathieu.Segond@th.u-psud.fr}}

\abstract{We have shown that  the lowest order QCD amplitude, i.e. the quark exchange
contribution, to the forward production  of a pair of longitudinally polarized
$\rho$ mesons in the scattering of two virtual photons
$\gamma^*(Q_1) \gamma^*(Q_2) \to \rho^0_L \rho^0_L$ factorizes in two 
different ways: the part with  transverse photons is described by the QCD factorization formula
involving the generalized distribution amplitude of two final $\rho$
mesons, whereas the part with longitudinally polarized photons takes
the QCD factorized form with the
 $\gamma^*_L \to \rho^0_L$ transition distribution amplitude.
Perturbative expressions for these, in general,
non-perturbative functions are obtained in terms of the $\rho-$meson distribution amplitude.}

\FullConference{International Workshop on Diffraction in High-Energy Physics -DIFFRACTION 2006 -\\
		 September 5-10 2006\\
		 Adamantas, Milos island, Greece}

\begin{document}

\section{Introduction}
\begin{figure}[htb]
\psfrag{r1}[cc][cc]{$\quad\rho(k_1)$}
\psfrag{r2}[cc][cc]{$\quad\rho(k_2)$}
\psfrag{p1}[cc][cc]{$\slashchar{p}_1$}
\psfrag{p2}[cc][cc]{$\slashchar{p}_2$}
\psfrag{q1}[cc][cc]{$q_1$}
\psfrag{q2}[cc][cc]{$q_2$}
\psfrag{Da}[cc][cc]{DA}
\psfrag{HDA}[cc][cc]{$M_H$}
\psfrag{M}[cc][cc]{$M$}
\centerline{\scalebox{0.9}
{
$\begin{array}{cccc}
\raisebox{-0.44 \totalheight}{\epsfig{file=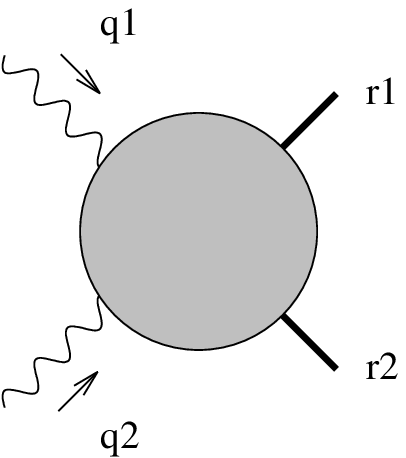,width=\scalingl}}&=&
\raisebox{-0.44\totalheight}{\epsfig{file=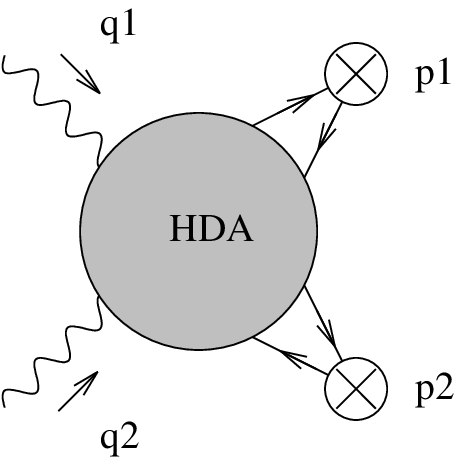,width=\scalingl}}
& \begin{array}{c}
\raisebox{0.4 \totalheight}
{\psfrag{r}[cc][cc]{$\quad\rho(k_1)$}
\psfrag{pf}[cc][cc]{$\slashchar{p}_2$}
\epsfig{file=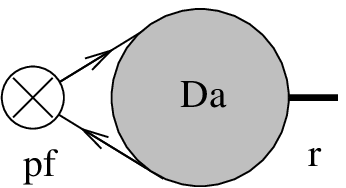,width=\scalingl}}\\
\raisebox{0.1 \totalheight}
{\psfrag{r}[cc][cc]{$\quad\rho(k_2)$}
\psfrag{pf}[cc][cc]{$\slashchar{p}_1$}
\epsfig{file=DA.eps,width=\scalingl}}
\end{array}
\end{array}
$}}
\caption{\small The amplitude of the process $\gamma^*(Q_1) \gamma^*(Q_2) \to \rho^0_L (k_{1})\rho^0_L(k_{2})$ in
the collinear factorization, in which we show the $M_H$ pertubative amplitude and the non-pertubative DA of $\rho$-mesons. \label{M}}
\end{figure}

The exclusive reaction 
\beq
 \label{proces}
  \gamma^*(q_1)
\gamma^*(q_2) \to \rho^0_L (k_{1})\rho^0_L(k_{2}) \ee  is a beautiful laboratory to explore the application of perturbative QCD and in particular the factorization theorems which decouple  large distance from short distance phenomena. We have calculated \cite{PSSW} the Born order of this process and demonstrated how the factorization properties of the amplitude emerge.  At higher energies, the gluon exchange contribution  dominates, both at the
Born order \cite{PSW} and in the resummed BFKL approach \cite{EPSW}. The Born order contribution with quark exchanges
is described by the
 same set of diagrams  which contributes to the  scattering of real photons
producing pions at large
momentum transfer, studied long ago \cite{BLphysrev24} in the framework of the factorized form of exclusive processes at fixed angle. In these processes, mesons are
described by their light-cone distribution amplitudes (DAs), as  illustrated in Fig.~\ref{M}.
Our present study can be seen as a complement of Ref.
\cite{BLphysrev24} for the case of the scattering with virtual
photons, i.e. with transverse and longitudinal polarizations, in the
 forward kinematics (see also Ref. \cite{DFKV}).
The virtualities $Q^2_i=-q^2_i$, $i=1,2$, supply the hard scale to
the process (\ref{proces}) which justifies the use of the QCD
collinear  factorization methods and the description of $\rho$
mesons by means of their distribution amplitudes.

First, we calculate, within this scheme, the scattering
amplitude of (\ref{proces}) at Born level. Then, we turn to the study
of two particular kinematical regions and show how the amplitude with transverse photons factorizes in a hard subprocess and a Generalized 
Distribution Amplitude (GDA) \cite{DM,GDA}, while on the other hand the amplitude with longitudinal photons
factorizes in a hard subprocess and a $\gamma^* \to \rho$ Transition Distribution
Amplitude (TDA) \cite{TDA}.

\section{The Born order amplitude}
\label{secborn}

The kinematics is quite simple in the forward regime that we are investigating. We use a Sudakov
decomposition with two lightlike vectors $p_1$ and $p_2$ with $2 p_1 . p_2 = s$ and write the photon and the meson momenta respectively as
\beq
q_{1} = p_1 - \frac{Q_{1}^2}{s} p_2 \;~~,~~  q_{2} = p_2 - \frac{Q_{2}^2}{s} p_1\;~~~~ \mbox{and} ~~~~k_{1} =  \left(1  - \frac{Q_{2}^2}{s} \right) p_1 \;~~,~~ k_{2} = \left(1  - \frac{Q_{1}^2}{s}\right) p_2\;.
\ee
The energy's positivity of produced $\rho'$s requires that $s \geq Q_i^2$. The squared invariant mass of the two $\rho-$mesons $W^2$ and the minimum squared momentum transfer are 
\beq
\label{Ws} 
W^2 = (q_{1}+ q_{2})^2 =s  \left(1  - \frac{Q_{1}^2}{s}\right)  \left(1  - \frac{Q_{2}^2}{s}\right)\;~~~\mbox{and}~~~ t_{min} = (q_{2}- k_{2})^2 = - \frac{Q_{1}^2 Q_{2}^2}{s}\;.
\ee

Contrarily to the case studied in Ref.\cite{BLphysrev24,DFKV }, notice that $t_{min}$ may not be large with respect to $\Lambda^2_{QCD}$ , depending on the respective values of
$Q_1^2$, $Q_2^2$ and $W^2$.


The scattering amplitude ${\cal A}$ of the process (\ref{proces}) can be written in the form
${\cal A}= T^{\mu\, \nu}\epsilon_\mu(q_1)\epsilon_\nu(q_2)\;,$
where the tensor $T^{\mu\, \nu}$ has in the above kinematics a simple decomposition 
\beq
\label{T}
T^{\mu\, \nu} = \frac{1}{2}g^{\mu\,\nu}_T\;(T^{\alpha\, \beta}g_{T\,\alpha\,\beta}) +
(p_1^\mu +\frac{Q_1^2}{s}p_2^\mu)(p_2^\nu
+ \frac{Q_2^2}{s}p_1^\nu)\;\frac{4}{s^2}(T^{\alpha\, \beta}p_{2\,\alpha}\, p_{1\,\beta})\;,
\ee
and where $g^{\mu\,\nu}_T=g^{\mu \nu} - (p_1^\mu p_2^\nu + p_1^\nu p_2^\mu)/(p_1 . p_2)$.
The longitudinally polarized
$\rho^0-$meson DA $\phi(z)$ is defined by the standard non-local correlator of quark anti-quark fields on the light cone.

The Born order contribution to the amplitude is calculated in the Feynman gauge
in a similar way as in the classical work of Brodsky-Lepage \cite{BLphysrev24} but in very different kinematics.
In our case the virtualities of photons supply the hard scale,
not the transverse momentum transfer.

The scalar components of the scattering amplitude (\ref{T}) read :
\begin{eqnarray}
\label{Ttr}
T^{\alpha\, \beta}g_{T\,\alpha \,\beta}&=& -\frac{e^2(Q_u^2 +Q_d^2)\,g^2\,C_F\,f_\rho^2}{4\,N_c\,s}
\int\limits_0^1\,dz_1\,dz_2\,\phi(z_1)\,\phi(z_2)
 \\
&&\hspace{-2cm}\times \left\{2\left(1-\frac{Q^2_2}{s}\right)\left(1-\frac{Q^2_1}{s}\right)\left[
\frac{1}{(z_2+\bar z_2\frac{Q_1^2}{s})^2(z_1+\bar z_1\frac{Q_2^2}{s} )^2} +
\frac{1}{(\bar z_2+ z_2\frac{Q_1^2}{s})^2(\bar z_1+ z_1\frac{Q_2^2}{s} )^2}
 \right] \; + \right.
\nonumber \\
 && \hspace{-3cm}\left.
\left(\frac{1}{\bar z_2\,z_1}- \frac{1}{\bar z_1\,z_2}  \right)
\left[ \frac{1}{1-\frac{Q^2_2}{s}}\left( \frac{1}{\bar z_2+ z_2\frac{Q_1^2}{s}}
                                      -\frac{1}{ z_2+ \bar z_2\frac{Q_1^2}{s}}   \right)
 - \frac{1}{1-\frac{Q^2_1}{s}}\left(\frac{1}{\bar z_1+ z_1\frac{Q_2^2}{s}}                                    - \frac{1}{ z_1+ \bar z_1\frac{Q_2^2}{s}}   \right) \right]
\right\} \nonumber \\
\label{Tlong}
T^{\alpha\, \beta}p_{2\,\alpha} \,p_{1\,\beta}&=& -\frac{s^2 f_\rho^2 C_F e^2 g^2(Q_u^2 +Q_d^2)}{8N_{c}Q_1^2 Q_2^2}
\int\limits_0^1\,dz_1\,dz_2\,\phi(z_1)\,\phi(z_2)
 \\
&&\times \left\{\frac{(1-\frac{Q^2_1}{s})(1-\frac{Q^2_2}{s})}{(z_1+\bar z_1 \frac{Q^2_2}{s})(z_2+\bar z_2\frac{Q^2_1}{s})}
+ \frac{(1-\frac{Q^2_1}{s})(1-\frac{Q^2_2}{s})}{(\bar z_1+ z_1 \frac{Q^2_2}{s})(\bar z_2+ z_2\frac{Q^2_1}{s})}
+ \frac{1}{z_2 \bar z_1} + \frac{1}{z_1 \bar z_2} \right\} \nonumber\;,
\end{eqnarray}

\noindent where $Q_u=2/3$ ($Q_d=-1/3$) denote the charge of the quark $u$ ($d$),
$C_F=(N_c^2-1)/(2 N_c)$ and $N_c=3.$ All integrals (\ref{Ttr})-(\ref{Tlong}) over quarks momentum fractions  $z_i$ are convergent. 

\section{$\gamma_T^* \gamma_T^* \to \rho^0_L \rho^0_L$ in the generalized Bjorken limit}
\label{secgda}

\begin{figure}[htb]
\psfrag{r1}[cc][cc]{$\quad\rho(k_1)$}
\psfrag{r2}[cc][cc]{$\quad\rho(k_2)$}
\psfrag{p1}[cc][cc]{$\slashchar{p}_1$}
\psfrag{p2}[cc][cc]{$\slashchar{p}_2$}
\psfrag{p}[cc][cc]{$\qquad\slashchar{P}\qquad \slashchar{n}$}
\psfrag{n}[cc][cc]{}
\psfrag{q1}[cc][cc]{$q_1$}
\psfrag{q2}[cc][cc]{$q_2$}
\psfrag{GDA}[cc][cc]{$GDA_H$}
\psfrag{Da}[cc][cc]{DA}
\psfrag{HDA}[cc][cc]{$M_H$}
\psfrag{M}[cc][cc]{$M$}
\psfrag{Th}[cc][cc]{$T_H$}
\centerline{\scalebox{1}
{$\begin{array}{cccccc}
\!\!\raisebox{-0.44\totalheight}{\epsfig{file=HDA.eps,width=\scalingD}}& \begin{array}{c}
\raisebox{0.51 \totalheight}
{\psfrag{r}[cc][cc]{$\quad\rho(k_1)$}
\psfrag{pf}[cc][cc]{$\slashchar{p}_2$}
\epsfig{file=DA.eps,width=\scalingA}}\\
\raisebox{0. \totalheight}
{\psfrag{r}[cc][cc]{$\quad\rho(k_2)$}
\psfrag{pf}[cc][cc]{$\slashchar{p}_1$}
\epsfig{file=DA.eps,width=\scalingA}}
\end{array}&=& \!
\raisebox{-0.44 \totalheight}
{\psfrag{r}[cc][cc]{$\quad\rho(k_1)$}
\psfrag{pf}[cc][cc]{$\slashchar{p}_2$}
\epsfig{file=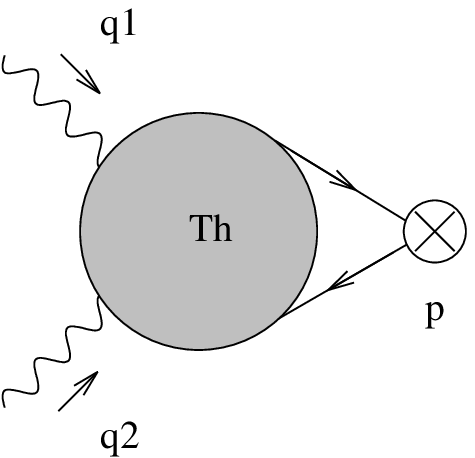,width=\scalingB}}&
\raisebox{-0.43\totalheight}{\epsfig{file=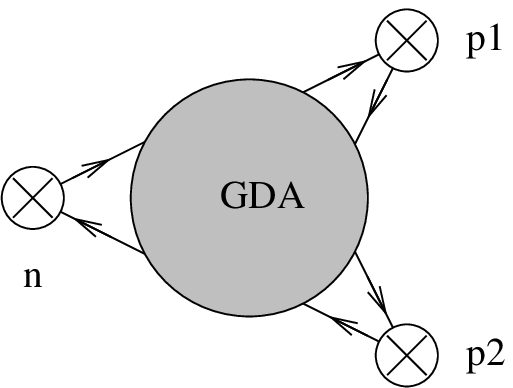,width=\scalingC}}& \begin{array}{c}
\raisebox{0.51 \totalheight}
{\psfrag{r}[cc][cc]{$\quad\rho(k_1)$}
\psfrag{pf}[cc][cc]{$\slashchar{p}_2$}
\epsfig{file=DA.eps,width=\scalingA}}\\
\raisebox{-0.0 \totalheight}
{\psfrag{r}[cc][cc]{$\quad\rho(k_2)$}
\psfrag{pf}[cc][cc]{$\slashchar{p}_1$}
\epsfig{file=DA.eps,width=\scalingA}}
\end{array}
\end{array}
$}}
\caption{Factorization of the  amplitude in terms of a GDA which is expressed in a pertubatively computed $GDA_H$ convoluted with the DAs of the $\rho$-mesons. \label{FactGDA}}
\end{figure}

\noindent In the region where the scattering energy $W$ is
small compared to the highest photon virtuality $Q_1$ \be
\label{scaling} \frac{W^2}{Q_1^2}=
\frac{s}{Q_1^2}\left(1-\frac{Q_1^2}{s}
\right)\left(1-\frac{Q_2^2}{s}  \right) \approx 1-\frac{Q_1^2}{s}\ll
1\;, \ee which leads to  kinematical conditions very close to the
ones considered in Ref.\cite{DM,GDA} for the  description of
$\gamma^* \gamma \to \pi \pi$ near threshold. In Ref.\cite{DM, GDA}
it was shown that, with initial transversally polarized photons, the
scattering amplitude factorizes at leading twist as the convolution
of a perturbatively calculable coefficient function  and a
generalized distribution amplitude (GDA). We  recover a similar
type of factorization with a GDA of the expression (\ref{Ttr}) also
in the process of our case (\ref{proces}), as illustrated in
Fig.~\ref{FactGDA}.

\noindent We first assume that $Q_1$ and $Q_2$ are not parametrically close, i.e. $1-Q_1^2/s \ll 1-Q_2^2/s\,.$

The leading contribution is given by the very last term in (\ref{Ttr}), 
\begin{eqnarray}
\label{Ttrscaling}
T^{\alpha\, \beta} \! g_{T\,\alpha \,\beta} \!  \approx \! \frac{e^2(Q_u^2 +Q_d^2)\,g^2\,C_F\,f_\rho^2}{4\,N_c\,W^2} \!
\!\int\limits_0^1\!\!dz_1\!dz_2 \phi(z_1)\!\phi(z_2)\!
\left(\!\frac{1}{\bar z_2 z_1}\!- \!\frac{1}{\bar z_1 z_2}\!  \right) \!  \! 
\left(\!\frac{1}{\!\bar z_1\!+ \!z_1\!\frac{Q_2^2}{s}}
 \!-\! \frac{1}{\! z_1\!+\! \bar z_1\!\frac{Q_2^2}{s}}   \right)\!.\;
\end{eqnarray}
To interpret this result in a factorized formula, let us first recall the definition of the leading twist GDA for a $\rho^0_L$ pair. We introduce the
vector $P=k_1+k_2 \approx p_1$, whereas the field coordinates  are the ray-vectors along
the light-cone direction $n^\mu=p_2^\mu/(p_1.p_2)$.
In our kinematics the variable $\zeta = (k_1n)/(Pn)$ characterizing the GDA equals $\zeta \approx 1$.
Thus we define the GDA of the $\rho^0_L$ pair $\Phi_q(z,\zeta,W^2)$ for $q=u,d$  by
the formula
\begin{eqnarray}
\label{GDA}
\langle \rho^0_L(k_1)\,\rho^0_L(k_2)| \bar q( \!\! -\alpha n/2)\,\slashchar{n}\,\exp \!
 \!\left[ \!\int\limits_{-\frac{\alpha}{2}}^{\frac{\alpha}{2}} \! \!dy \, n_\nu\, A^\nu (y) \! \right]\!\!
q(\alpha \;n/2)|0\rangle
\nonumber 
=
 \int\limits_0^1\!\!dz\,e^{-i(2z-1)\alpha (nP)/2}\Phi^{\rho_L\rho_L}(z,\zeta,W^2)\;.
\end{eqnarray}

\noindent We can then calculate the GDA  $\Phi_q(z,\zeta,W^2)$ in the Born order of the perturbation theory.
The gluonic Wilson line does not give any contribution in our particular kinematics. The  remaining contributions to
the correlator at order $g^2$  
lead to the result
\be
\label{Phi}
\Phi^{\rho_L\rho_L}(z,\zeta\approx 1,W^2)
= -\frac{f_\rho^2\,g^2\,C_F}{2\,N_c\,W^2}\int\limits_0^1\,dz_2\,\phi(z)\,\phi(z_2)
\left[\frac{1}{z\bar z_2} -\frac{1}{\bar z z_2} \right]\;.
\ee

\noindent In the case of a quark of a given flavour, the hard part $T_H$ of the amplitude equals
\begin{equation}
\label{hardGDA}
T_H(z) = -4\,e^2\,N_c\,Q_q^2\,\left( \frac{1}{\bar z +z\,\frac{Q_2^2}{s}} -
   \frac{1}{z + \bar z\,\frac{Q_2^2}{s}} \right) .
\end{equation}
Eqs.~(\ref{Phi}, \ref{hardGDA}) taken together with the flavour
structure of $\rho^0$ enables us to write (\ref{Ttrscaling}) as
\be
\label{GDAfinal}
T^{\alpha\, \beta}g_{T\,\alpha \,\beta}=
\frac{ e^2}{2}\left(Q_u^2 +Q_d^2  \right)\int\limits_0^1 \,dz\,\left(\frac{1}{\bar z +z \frac{Q_2^2}{s}}
- \frac{1}{ z +\bar z \frac{Q_2^2}{s}}  \right)\Phi^{\rho_L\rho_L}(z,\zeta\approx 1,W^2)\;,
\ee
which shows the factorization of $T^{\alpha\, \beta}g_{T\,\alpha \,\beta}$
 into the hard part and the GDA (up to corrections of order
$W^2/Q_1^2$) .
The Eq.~(\ref{GDAfinal})
 is  the limiting case for $\zeta \to 1$
of the original equation derived by D.~M{\"u}ller et al. \cite{DM}.  Let us finally note that in general GDAs are complex functions. In our case, i.e. in the Born approximation, the hard part (\ref{hardGDA})
and the GDA (\ref{Phi}) are real quantities. This is due to the use of the 
real DA of $\rho$-mesons and to the absence of the $s-$channel cut of 
hard diagrams in the Born approximation. 

\noindent The  peculiar case of parametrically close photon virtualities deserves particular attention.
In this case, there are subtle problems for defining the light-cone vector $P$  since the two outgoing mesons should be now treated in an almost symmetric way. In order to circumvent this difficulty,
it is useful to start from the factorized formula (\ref{GDAfinal}) in the kinematical domain (\ref{scaling}),
assuming $Q_1 > Q_2.$ 
Let us continue (\ref{GDAfinal}) in $Q_2$ up to $Q_2=Q_1.$ To control this continuation, one should restore the $Q_1^2$ and $Q_2^2$ dependence, encoded in $\zeta$  and $W^2$ through $W^2/\zeta= s (1-Q_1^2/s),$ as 
\be
\label{PhiExact}
\Phi^{\rho_L\rho_L}(z,\zeta,W^2)
= -\frac{f_\rho^2\,g^2\,C_F \, \zeta}{2\,N_c\,W^2}\int\limits_0^1\,dz_2\,\phi(z)\,\phi(z_2)
\left[\frac{1}{z\bar z_2} -\frac{1}{\bar z z_2} \right]\;.
\ee
The hard part in (\ref{GDAfinal}) is proportional to $1-Q_2^2/s.$  
This factor $1-Q_2^2/s, $ now of the order of $1-Q_1^2/s ,$ starts to play the role of a suppression factor. The amplitude (\ref{GDAfinal})
 which is proportional to $\frac{1}{s} \frac{1-Q_2^2/s}{1-Q_1^2/s},$ now behaves as $1/s \sim 1/Q^2.$
In the leading twist approximation, this factorized result vanishes. This observation is confirmed by the result (\ref{Ttr}) of the direct calculation. Indeed, in the case $Q_1=Q_2=Q,$ the magnifying 
factor $1/(1-Q^2/s)$ in the two terms of the last line of (\ref{Ttr})
is not present anymore. Thus, the resulting amplitude should be considered as a higher twist contribution.

\section{$\gamma_L^* \gamma_L^* \to \rho^0_L \rho^0_L$
in the generalized Bjorken limit}
\label{sectda}

In the regime $Q_{1}^2  \gg  Q_{2}^2 \; $, it  has been advocated \cite{TDA} that the amplitude with
initial longitudinally polarized  photons, should factorize as the
convolution of a perturbatively calculable coefficient function  and
a $\gamma \to \rho$ transition distribution amplitude (TDA) defined
from the non-local quark correlator
  \beq
\int \frac{dz^-}{2\pi} e^{-ixP^+z^-} \langle \rho(p_{2}) | \bar q(-z^-/2) \gamma^+ q(z^-/2)) | \;\gamma(q_{2}) \rangle \;,
\ee
 which shares many properties (including the QCD evolution equations) with the generalized parton
 distributions \cite{DM,GPD} succesfully introduced to describe deeply virtual Compton scattering.

The factorization properties of the scattering amplitude in this
domain are conventionally described by using the now standard
notations of GPDs. For that we rewrite the momenta of the particles
involved in the process as
\begin{eqnarray}
\label{GPDkin}
q_1\!=\!\frac{1}{1\!+\!\xi}n_1\! -\!2\xi\!n_2\;,\;\;\;\;k_1=\frac{1\!-\!\frac{Q_2^2}{s}}{1\!+\!\xi}\,n_1, \;\;\;
 q_2\!=\!-\frac{Q_2^2}{(1\!+\!\xi)s}n_1 + (1\!+\!\xi) n_2\;,\;\;\;\;k_2=(1\!-\!\xi)\,n_2\;,
\end{eqnarray}
where $\xi$ is the skewedness parameter which equals $\xi= Q_1^2/(2s - Q_1^2)$   and the new $n_i$ Sudakov light-cone
vectors are related to the $p_i$'s as $p_1=\frac{1}{1+\xi}\,n_1$ and $p_2=(1+\xi)\,n_2$
with
 $p_1 \cdot p_2=n_1 \cdot n_2=s/2$. We also introduce the
average ``target'' momentum $P=\frac{1}{2}(q_2 +k_2)$ and the momentum transfer $\Delta=k_2-q_2$. We still restrict our study to the strictly forward case with $t=t_{min} =-2\xi Q_2^2/(1+\xi)$.

Defining the new variable $x$ through   \ $z_2=(x-\xi)/(1-\xi)$ with $x \in [\xi,1]$,  the expression in Eq.~(\ref{Tlong})
 can be put  into the form:
 \begin{eqnarray}
\label{4terms}
&&\int\limits_0^1\,dz_1\,dz_2\,\phi(z_1)\,\phi(z_2)\left\{....\right\}=
\int\limits_{-1}^1\,dx\,
\int\limits_0^1\,dz_1\,\phi(z_1)\,\left( \frac{1}{\bar z_1(x-\xi)}+ \frac{1}{z_1(x+\xi)}  \right) \\
&&\times \, \left[\Theta(1\ge x \ge \xi) \phi\left(\frac{x-\xi}{1-\xi} \right) -
\Theta(-\xi \ge x \ge -1) \phi\left( \frac{1+x}{1-\xi} \right)
 \right]\;. \nonumber
\end{eqnarray}
 \vspace{-.5cm}
 \begin{figure}[htb]
\psfrag{q1}[cc][cc]{$q_1$}
\psfrag{q2}[cc][cc]{$q_2$}
\psfrag{p1}[cc][cc]{$\slashchar{p}_1$}
\psfrag{p2}[cc][cc]{$\slashchar{p}_2$}
\psfrag{n}[cc][cc]{$\slashchar{p}_1$}
\psfrag{p}[cc][cc]{$\slashchar{p}_2$}
\psfrag{Tda}[cc][cc]{$TDA_H$}
\psfrag{Th}[cc][cc]{$T_H$}
\psfrag{Da}[cc][cc]{DA}
\centerline{\scalebox{1.05}
{
$\begin{array}{c}
\begin{array}{cc}
\raisebox{-0.44 \totalheight}{\epsfig{file=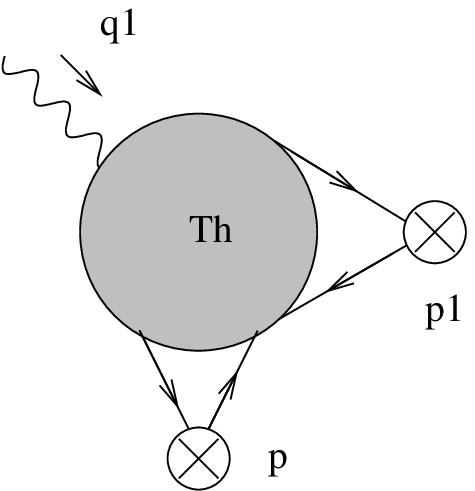,width=\scaling}}&
\raisebox{-0.25\totalheight}
{\psfrag{r}[cc][cc]{$\qquad \rho(k_1)$}
\psfrag{pf}[cc][cc]{\raisebox{-1.05 \totalheight}{$\slashchar{p}_2$}}\epsfig{file=DA.eps,width=\scalingm}}
\end{array}\\
\\
\begin{array}{cc}
\raisebox{-0.44 \totalheight}{\epsfig{file=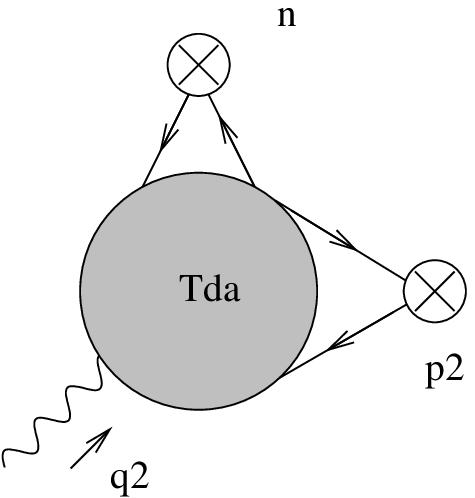,width=\scaling}}&
\raisebox{-0.54\totalheight}
{\psfrag{r}[cc][cc]{$\qquad \rho(k_2)$}
\psfrag{pf}[cc][cc]{\raisebox{-1.05 \totalheight}{$\slashchar{p}_1$}}\epsfig{file=DA.eps,width=\scalingm}}
\end{array}
\end{array}
$}}
\caption{Factorization of the amplitude in terms of a TDA (lower part) which is itself the convolution of a hard term $TDA_H$ and a DA of the $\rho$-meson.   \label{FactTDA}}
\end{figure}

\noindent To rewrite Eq.~(\ref{4terms}) in a form
corresponding to the QCD factorization with a TDA, as illustrated in
Fig.~\ref{FactTDA}, we first calculate the hard part $T_H(z_1,x)$ of the scattering amplitude, at order $g^2$, for a meson built from a quark with a single flavour:
\begin{eqnarray}
\label{hardpart}
\hspace{-1cm} T_H(z_1,x)\!&=\!&-i f_\rho g^2 e Q_q \frac{C_F \phi(z_1)}{2N_c Q_1^2}
\epsilon^\mu(q_1)\!\left(\!2\xi n_{2\mu} \! +\!\frac{1}{1\!+\!\xi}\!n_{1\mu}
\right)
\!\!\left[\!\frac{1}{z_1\!(x\!+\!\xi-\!i \epsilon)}\! +\! \frac{1}{\bar
z_1\!(x\!-\!\xi\!+\!i \epsilon)}    \right]\,
\end{eqnarray}
\noindent and obviously coincides with the hard part of the $\rho-$meson
electroproduction amplitude (\ref{4terms}). The tensorial structure
 $2\xi\,n^\mu_{2\,} +\frac{1}{1+\xi}n^\mu_{1}=p_1^\mu + Q_1^2/s\,p_2^\mu$
coincides with the one
present in Eq.(\ref{T}). 

Passing to the TDA, let us consider the definition of
$\gamma^*_L(q_2) \to \rho^q_L(k_2) $ TDA, $T(x,\xi,t_{min})$, in
which we assume that the meson is built from a quark with a single
flavour, $\rho^q_L(k_2)= \bar q q$. The vector $P=1/2(q_2
+k_2)\approx n_2$ in our kinematics, and the ray-vector of
coordinates is oriented along the light-cone vector
$n=n_1/(n_1.n_2)$. The computation of the TDA gives 
\begin{eqnarray}
\label{defTDA}
&&\int\,\frac{dz^-}{2\pi}\,e^{ix(P.z)}\,\langle \rho^q_L(k_2)|
\bar q(-z/2)\hat n\,e^{-ieQ_q\,\int\limits_{z/2}^{-z/2}\,
dy_\mu\,A^\mu(y) }q(z/2)|\gamma^*(q_2)\rangle
\nonumber \\
&&= \frac{e\,Q_q\,f_\rho}{P^+}\,\frac{2}{Q_2^2}\,
\epsilon_\nu(q_2)\left((1+\xi)n_2^\nu +\frac{Q_2^2}{s(1+\xi)}n_1^\nu\right)\,T(x,\xi,t_{min})\;,
\end{eqnarray}
in which we explicitly show the electromagnetic Wilson-line assuring
the abelian gauge invariance of the non-local operator. We omit the Wilson line
required by the non-abelian QCD invariance since it does not play
any role in this case. Note that the factor $(1+\xi)n_2^\nu
+\frac{Q_2^2}{s(1+\xi)}n_1^\nu= p_2^\nu + \frac{Q_2^2}{s}p_1^\nu$
corresponds to a part of the tensorial structure of the second term
in Eq.~(\ref{T}). The perturbative calculation of the matrix element in (\ref{defTDA}) leads to:
\begin{equation}
\label{tda}
T(x,\xi,t_{min})\equiv N_c \left[\Theta(1\ge x \ge \xi)\, \phi\left(\frac{x-\xi}{1-\xi} \right) -
\Theta(-\xi \ge x \ge -1) \, \phi\left( \frac{1+x}{1-\xi} \right) \right]\; .
\end{equation}
\noindent In particular, the contribution to the rhs of (\ref{defTDA})
proportional to the vector $n_1^\nu$ (or $p_1^\nu$) corresponds to
the contribution coming from the expansion of the electromagnetic
Wilson line.
\noindent Putting all factors together and restoring  the flavour structure of the $\rho^0$, we obtain the
 factorized form (up to corrections of order $Q_2^2/Q_1^2$) involving a TDA, of Eq.~(\ref{Tlong}) as
\begin{equation}
\label{tdafactor}
T^{\alpha\,\beta}p_{2\,\alpha}p_{1\,\beta}=
-if_\rho^2e^2(Q_u^2+Q_d^2)g^2\,\frac{C_F}{8N_c}\int\limits_{-1}^1 dx\,\int\limits_0^1 dz_1\,
\left[\frac{1}{\bar z_1(x-\xi)} + \frac{1}{z_1(x+\xi)} \right]\,T(x,\xi,t_{min})\;.
\end{equation}

\section{conclusion}
In conclusion, we have shown that the perturbative analysis of the process
$\gamma^*\;\gamma^*\to \rho^0_L \rho^0_L$ in the Born approximation
leads to two different types of QCD factorization which are dictated not only by the kinematics but also by the
polarization states of the photons. The arbitrariness in choosing values of photon virtualities shows that there may exist
an intersection region where both types of factorization are simultaneously valid.

\section*{Acknowledgments}

This  work is partially supported by the Polish Grant 1 P03B 028
28. L.Sz.\ is a Visiting Fellow
of the Fonds National pour la Recherche Scientifique (Belgium).


\end{document}